\documentclass{article}

\PassOptionsToPackage{numbers, compress}{natbib}

\usepackage[final]{neurips_2019_ml4ps}

\usepackage[utf8]{inputenc} 
\usepackage[T1]{fontenc}    
\usepackage{hyperref}       
\usepackage{url}            
\usepackage{booktabs}       
\usepackage{amsfonts}       
\usepackage{nicefrac}       
\usepackage{microtype}      
\usepackage{graphicx}
\usepackage{natbib}

\usepackage{color}
\usepackage{amsmath}

\title{Interpretability Study on Deep Learning for Jet Physics at the Large Hadron Collider}

\author{%
   Taoli Cheng \\
  Montreal Institute for Learning Algorithms (MILA)\\
  Université de Montréal\\
  \texttt{chengtaoli.1990@gmail.com} \\
}

\begin{document}

\maketitle

\begin{abstract}
Using deep neural networks for identifying physics objects at the Large Hadron Collider (LHC) has become a powerful alternative approach in recent years. After successful training of deep neural networks, examining the trained networks not only helps us understand the behaviour of neural networks, but also helps improve the performance of deep learning models through proper interpretation. We take jet tagging problem at the LHC as an example, using recursive neural networks as a starting point, aim at a thorough understanding of the behaviour of the physics-oriented DNNs and the information encoded in the embedding space. We make a comparative study on a series of different jet tagging tasks dominated by different underlying physics. Interesting observations on the latent space are obtained.
\end{abstract}

\section{Introduction}

Applying deep neural networks in jet tagging has been gaining a lot of progress during latest years. The community has reached the common sense that 1), DNNs taking in low-level features are able to perform well in jet tagging tasks and automatically carry out feature engineering. 2), different architectures have been explored (including image-based CNNs \cite{Cogan:2014oua, deOliveira:2015xxd} , momentum-vector-based RNNs \cite{Egan:2017ojy}, physics-inspired Recursive NNs \cite{Louppe:2017ipp}, graph models \cite{Henrion:DLPS2017, Moreno:2019bmu} and point clouds \cite{Qu:2019gqs}, etc.). These architectures perform equivalently well in regard of tagging efficiency, and the results support for each other \cite{Kasieczka:2019dbj}. Despite the fact that this is an encouraging message, it also reminds us that there might be information redundancy in architecture search. It will be interesting if we can extract the common part among different architecture models to reduce information redundancy and also combine uniquely learnt information from different models to enhance tagging performance. 3), After the establishment of well-performing jet tagging DNNs, the next step would be pushing this further by interacting with our physics theoretical framework. 

Previous attempts to visualize and interpret DNNs (\cite{deOliveira:2015xxd,Komiske:2018cqr}) have been focusing on general DNNs activation and filter visualization. This gives us some intuitive sense about what information has been learnt. However, DNNs interpreted in this way is not sufficient for robust physics analysis. It has been obscure regarding finding a clear path between DNNs and physics-theory observable system.

In this work, we are trying to construct the connection with a series of studies varying from latent space probing to tailored saliency study.
We use different jet tagging tasks to form a comparative study, including W/QCD, Top/QCD, quark/gluon classifications.
In these classification tasks, different factors dominate since the causal factor resides in different part of the jet clustering trees. We know that the very essence of object identification in High Energy Physics is to "see" underlying quantum properties which evolves into low energy as observed patterns in the detector space. And our ultimate hope is in the same vein, i.e. to probe into the underlying physics structure, but not only detecting objects.
As for DNN architectures, we take advantage of the tree structured Recursive Neural Networks (RecNNs) \cite{Louppe:2017ipp} that utilize the well-established jet clustering algorithms, which encoded 
the radiation patterns within the clustering history. 
A jet representation is obtained through embedding each clustering node recursively along the clustering tree, and can then be easily fed into downstream tasks.
This approach produces interesting interpretation of encoded information within neural nets and also provides physics-friendly method for bridging physics theory and DNN-interpretation.

Although all the interpretation studies here are carried with the architecture of tree-based RecNNs and interpreted in the scope of supervised classification problems, the proposed methods and findings can be adaptively applied to other architectures and unsupervised learning cases.
As the last step within the interpretation loop, how to utilize the information gained by analyzing latent space, relevance and sensitivities to improve DNN performance and clarify other relevant aspects such as uncertainties and robustness, would be worth further exploring.

We use similar training datasets and neural network architecture along with parameters as in \cite{Cheng:2017rdo, Louppe:2017ipp} throughout this work. For all the tasks, we constrain jet transverse momentum to be around 600 GeV for a more meaningful comparison across tasks.
We feed input features ($p_i, \eta_i, \phi_i, E_i, E_i/E_J, p_{Ti},  \theta_i=2 \arctan(\exp{(-\eta_i)})$) for every jet constituent, into the jet clustering history to recursively build up the jet embedding. 
A more detailed description on the datasets and neural network architecture can be found in Appendix \ref{app:data}.

\section{Probing Jet Embedding Space}
\label{sec:probing}

Here we employ the ``probing method'' \cite{2016arXiv161001644A} to discover what has been presented in the latent space. By feeding the learnt latent embedding representations into auxiliary probing tasks, one will be able to, in a most straight-forward way, find the relevance between latent space and our target information, and thus get a hint on how well the latent space is aligned with the space of specific physics observables.

 After being trained on the input space of particle four momentum $\{(E, P_x, P_y, P_z)_i\}$ (or other variants such as $(p_T, \eta, \phi, m)_i$ of the input vector) in classification tasks, the embedding vectors of jets $\{h^J\}$ are then taken to be investigated in a task-independent way using linear classification or regression to link with our physics observable space $\{O_i\}$.
To be more specific, we take the \emph{Generalized Angularites} $\lambda_\beta^\kappa$ \cite{Larkoski:2014pca} as a systematic summary of common jet observables, as expressed in Eqn. \ref{eq:angularities}, with $z_i$ being the transverse momentum fraction of i-th jet component and 
$\theta = \frac{R_i}{R}$, where $R_i$ is the azimuthal distance to the jet axis and R is the jet radius.

\begin{equation}
    \lambda_\beta^\kappa = \sum_{i \in Jet} z_i^\kappa \theta_i^\beta
    \label{eq:angularities}
\end{equation}

We will then have in the ($\kappa$, $\beta$) space, for instance, jet mass as (1,2). We report, in Table \ref{tab:linear_probing}, the returned R2 score of the linear regression for jet angularities and jet N-subjettiness \cite{Thaler:2010tr} ($\tau_N$) which is very useful in probing ``prongness'' substructure of jets. R2 score is defined as the ratio of explained variance and total variance, with $R2=1$ corresponding to perfect fitting.

\begin{table}[]
    \centering
    \begin{tabular}{c|c|c|c}
    \toprule
    Jet Observables & R2(W/QCD) & R2(Top/QCD) & R2(q/g) \\ 
    \midrule
    multiplicity ($\lambda_0^0$) &  \textbf{0.68} & 0.55  & \textbf{0.75}\\
    width ($\lambda_1^1$) & \textbf{0.74} & \textbf{0.82}  &  0.55 \\
    $M_j$ ($\lambda_2^1$)   & \textbf{0.77} & \textbf{0.80} & 0.55\\ 
    $p_T^D$ ($\lambda^2_0$) & 0.30 & 0.36 & \textbf{0.65}\\
    \midrule
    $\tau_1$ & 0.67 & \textbf{0.78} & 0.54 \\
    $\tau_2$ & 0.67 &  0.62 &  0.56 \\
    $\tau_3$ & 0.67 & 0.53 & \textbf{0.61} \\
    $\tau_{2/1}$  & 0.39 & 0.35 & 0.20 \\ 
    $\tau_{3/2}$  & 0.16 & 0.39 & 0.09\\ 
    \bottomrule
    \end{tabular}
    \caption{R2 scores in linear regression for the latent jet representation, displayed for different classification tasks (W/QCD, Top/QCD, quark/gluon). Three observables with highest R2 scores are highlighted for each classification task.}
    \label{tab:linear_probing}
\end{table}

From Table \ref{tab:linear_probing}, one can see that among three classification tasks, W/QCD and Top/QCD both give high probing score on jet width ($\lambda_1^1$) \cite{Gallicchio:2012ez} and jet mass $M_j$ ($\lambda_2^1$), while quark/gluon gives high score on jet multiplicity ($\lambda_0^0$) and $p_T^D$ ($\lambda^2_0$) \cite{Chatrchyan:2012sn}. This is interesting, because in ($\kappa$, $\beta$) space, 
(1,1) and (1,2) are both IRC safe observables,
while the other two IRC unsafe angularities give important information in quark/gluon discrimination \cite{Larkoski:2014pca}.
Given the previous study results on $\lambda_\beta^\kappa$ quark/gluon discrimination using mutual information in \cite{Larkoski:2014pca}, it's foreseeable that further study on the full angularity space or using more sophisticated probing methods perhaps will lead to very interesting observations.
Another interesting point is that the ratios of N-subjettiness ($\tau_{2/1},~\tau_{3/2}$), which play very important role in traditional prongness jet substructure tagging including both W tagging and Top tagging, are not strongly manifested in the embedding representation. Rather, the N-subjettiness ($\tau_1,~\tau_2,~\tau_3$) themselves are much more directly related with the latent representation, indicating a more fundamental role in the latent space under investigation. 
Beside the characteristic jet observables studied above,
the authors in \cite{Datta:2019ndh, Larkoski:2019nwj} attempted to find the best discriminative observables, constructed from the N-subjettiness basis, for jet tagging problems with a machine learning approach. It will thus also be very interesting to see how these new observables are related to the latent embeddings. 

Beside embedded jet representation, we would like to see how the representation changes along the clustering trees. This is intrinsically difficult since every jet has different clustering history. As a naive attempt, we calculated the mass direction and want to see how the mass direction shift when passing it down the clustering tree. We observed that sometimes when the mass direction changes drastically (i.e. the mass direction calculated with jet embeddings doesn't fit with downwards clustering nodes), that's close to where the hard splitting (resonance decay) happens. However, a more systematic investigation will be necessary to lead to a confident conclusion.

\paragraph{Embedding Transferability}
As we have seen that some general physics observables (or relevant information) can be induced in the latent space by classification tasks, we expect the latent representation learnt by a specific classification task is informative enough to apply to other physics processes, where some common features may be transferable.
We try to pass the jet embeddings learnt by one classification task to another classification problem, to see how general or transferable the learnt jet representations are. Results show that for similar tasks (W/QCD and Top/QCD), the embeddings transfer very good performance, while the transferability decreases a bit for different types of tasks (such as across W/QCD and quark/gluon). More details can be found in Appendix \ref{app:transfer}.

\paragraph{Other Network Architectures}
We have shown linear probing for RecNNs. In Appendix \ref{app:models}, we also present linear probing results (only in Top/QCD classification) for other important architectures such as FCN, LSTM and CNN, with only low-level features as input. For some of these models, there is no clear 
\emph{Embedding Layer}, so we investigated all the 3 latest layers. R2 scores from these architectures are not as impressive as in RecNN, but still some common trends can be observed. For all these architectures, jet width, jet mass and $\tau_1$ ($\tau_2$) always dominate. Due to the image-based nature of CNN, the hidden representation is always in the pixel space. This makes it difficult to directly link with physics observables. So in the third latest layer \emph{Hidden(-3)} (the flatten layer), R2s are always close to 0. In later dense layers, R2 increases a bit, indicating some abstract features are learnt here. Comparatively, RecNN obviously surpass other models in learning physics observables in latent embeddings. While these analyses are only using a very simple linear regression, we expect that more complex analysis will also bring interesting observations.

\section{Interpreting on Jet Lund Plane}
\label{sec:lund_plane}
 
 Gradient-based saliency maps \cite{DBLP:journals/corr/SimonyanVZ13} give us a general sense on how input space affects activation in neural networks. However, this method is not straight-forward in revealing underlying physics. For jet physics, the most important building block is jet splitting mechanism, which tells us the underlying dominating interactions. Taking advantage of the base clustering structures within RecNNs, we can peek into every splitting easily, even within a neural network setup.
 In order to explore the jet splitting mechanisms and corresponding neural network activations, we map the saliency onto \emph{Jet Lund Planes}, giving a meaningful, in a physics sense, visualization.

Lund diagrams \cite{Dasgupta:2013ihk,Dreyer:2018nbf} build theoretically useful framework for jet splitting by expressing emissions on the lund plane $(\Delta, k_t)$, where $\Delta$ denotes angular distance between two splitting branches and $k_t$ denotes the relative transverse momentum of the softer branch (the emission). It gives a handy description on soft and collinear emissions within jets. And it separately display different kinematic regions with clear underlying physics. Soft-collinear emissions are emitted uniformly on lund plane and the region where hard splitting happens can always be easily spot. This gives us a navigation map to identify the physics nature of most sensitive nodes within neural networks.

\paragraph{Method}
We display gradient-based saliency maps for different jet classification tasks comparatively. The saliency for Recursive Neural Networks \cite{Louppe:2017ipp, gilles_github} is defined, in Eqn. \ref{eq:saliency}, as the $l_2$ norm of relative gradient sensitivity passed from the classifier output $L$ of every clustering embedding node $h_i$ normalized with respect to $l_2$ norm of the gradient sensitivity of final jet embedding node $\frac{\partial L}{\partial h^J}$.
\begin{equation}
 \left \lVert \widehat {\frac{\partial L}{\partial h_i}} \right \rVert_2 
 = \frac{\left \lVert {\frac{\partial L}{\partial h_i}} \right \rVert_2}{\left \lVert {\frac{\partial L}{\partial h^J}} \right \rVert_2}
 \label{eq:saliency}
\end{equation}
Saliency maps can be examined with in the tree-structure itself, however, only in an individual-based way, again since every jet has its own clustering history.
In order to have a collective impression on the sensitivity distribution for all the clustering nodes without losing track of underlying physics manifestation, we map the saliency on to jet lund plane to compare across tasks and across different kinematic regions dominated by different splitting mechanisms.

\begin{figure}[h]
    \centering
    \includegraphics[width=0.3\textwidth]{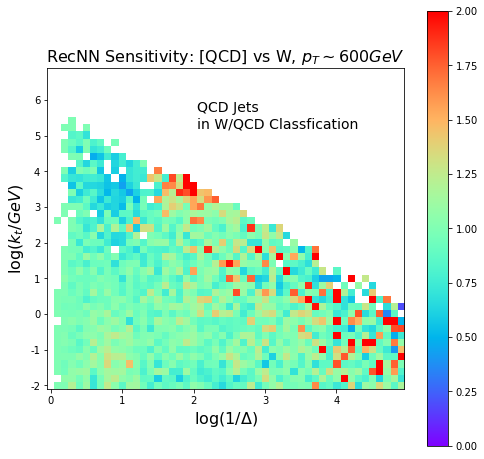}
    \includegraphics[width=0.3\textwidth]{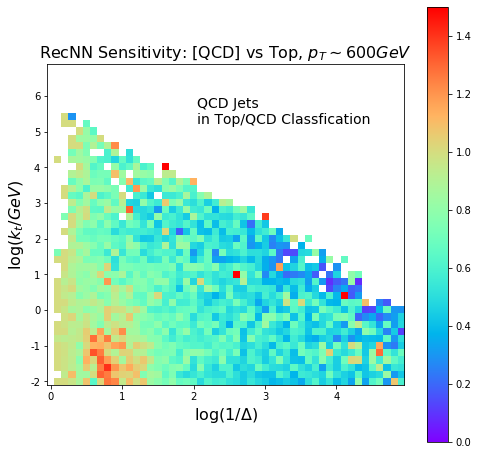}
    \includegraphics[width=0.3\textwidth]{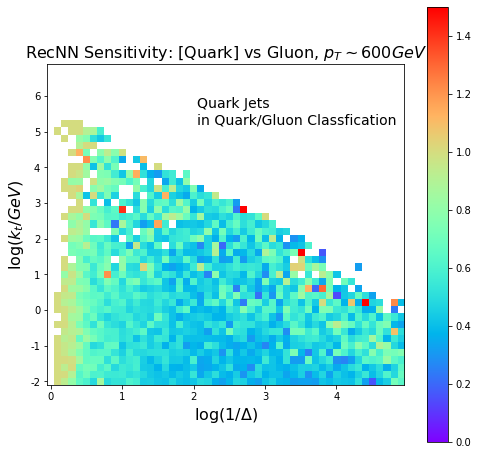} \\
    
    \includegraphics[width=0.3\textwidth]{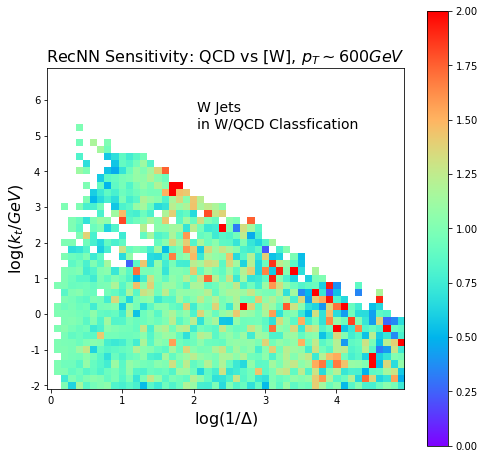}
    \includegraphics[width=0.3\textwidth]{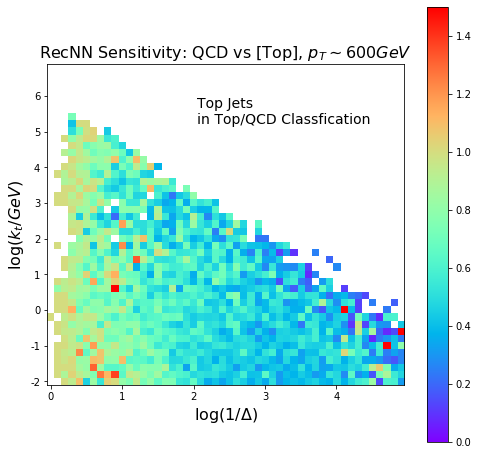}
    \includegraphics[width=0.3\textwidth]{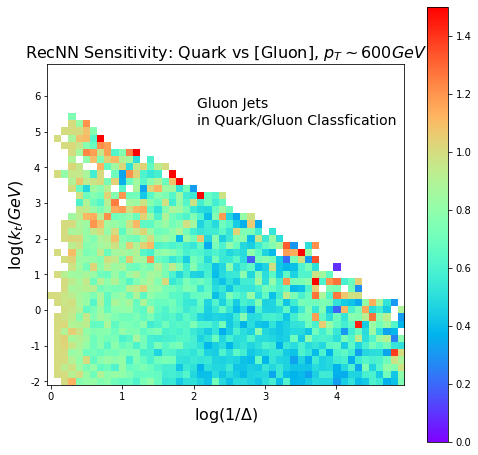}
    \caption{Saliency sensitivity mapped onto Jet Lund Plane. \textbf{Left}: QCD and W jets in W/QCD Classification; \textbf{Middle}: QCD and Top jets in Top/QCD Classification; \textbf{Right}: Quark and Gluon jets in Quark/Gluon Classification.}
    \label{fig:saliency}
\end{figure}

\paragraph{Results}
We present the averaged saliency sensitivity in Fig. \ref{fig:saliency}. For W/QCD, Top/QCD and quark/gluon classification, we show the saliency map on Lund Plane for both classes. This facilitates two purposes: class saliency within tasks and same-class saliency comparison across tasks. To make the illustration clearer, we restrict the saliency intensity (overflows are mapped onto upper bound) to be within [0, 2] for W/QCD and [0, 1.5] for the other two classifications. More plots for uniform range, which make easier comparison, are included in Appendix \ref{app:lund}.

From Fig. \ref{fig:saliency}, one can observe that:
\begin{itemize}
    \item For the three different tasks, the sensitivity maps on the lund plane give very different characteristic patterns. For W/QCD, the sensitivity is generally higher than other two tasks.
    \item From the left column for W/QCD classification, there are more activities along the diagonal boundary. This is where hard-collinear splitting happens. Notice that even for QCD jets, there is a high-sensitivity spot close to location (2, 3.5), which corresponds to hard splitting at a mass between 60 GeV (which is the mass peak of QCD jet samples) and 80 GeV (W boson mass). This might be related to the two-prongness (hard splitting), where the characteristics of W jets shapes the sensitivity activities within the comparing QCD jets.
    It's not clear if some high activity in collinear region is related to IRC safety robustness of the network. (IRC safety study in \cite{Louppe:2017ipp} showed that for ``collinear10-max'' (applies collinear splits to the 10 highest $p_T$ jet constituents), background rejection decreases a bit.)

    \item  From W/QCD to Top/QCD , the relative sensitive regions of QCD jets change drastically.  The left boundary corresponds to the final clustering step which in C/A will have large angles. In Top/QCD classification, sensitivity activity slightly decreases along the $\log(1/\Delta)$ axis. (the red spot of lower-left corner in QCD jets is curious though.) Compared to W/QCD, the hard-collinear region is much less sensitive.
     This may be partly because top jets with heavier mass have generally more decay constituents and correspondingly longer cascading chain and more complex jet substructure, where saliency passing down longer chains ending in collinear splittings tends to decrease or vanish. The top mass hard-splitting regions in the upper part of the triangle is less obvious compared to W/QCD case, but still relatively important in Top jets.
    \item Finally for quark/gluon classification, it seems that large-angle and some hard-collinear splittings (especially for gluon jets) are relatively important as shown in the plots. Generally speaking, saliency will effectively run through the whole clustering tree in quark/gluon tagging, where multiplicity plays an important role. So we don't see much sensitivity decrease along the $\log(1/\Delta)$ axis, while saliency passing from large angle down to small angles.
\end{itemize}

\section{Conclusions and Extended Studies}
\label{sec:conc}

We have shown some interesting results regarding probing the hidden representations learnt by jet tagging oriented deep neural networks, taking tree-structured recursive neural networks as an example. A cross-task comparative study helps us with revealing and understanding the information encoded in DNNs. In order to interpret the sensitivity within learnt neural net models in a physical way, we combine saliency tree maps and jet lund plane, and presented the mapped sensitivity across different jet classification tasks. Results show that when feeding in only low-level features of jet constituents, there is still high relevance with general physics observables in the latent embeddings; and the saliency maps give very characteristic patterns depending on the classification tasks. 

we only presented results for RecNNs. Some of the methods are easier to explore within RecNNs because of its physics-inspired architecture. In general, one can probe the embedding space or hidden layers to gain some basic information. 
However, for instance, it will be very difficult for CNNs to interpret in Lund Plane, since pixelization of jet images washes out the individual particle information. We expect some more suitable methods can be found for these architectures. On the other hand, neural networks directly featured with Lund coordinates might be another playground for Lund Plane interpretability study. And a systematic study compares different neural network architectures across different classification tasks will give more evident information on the model interpretability.
Depending on the specific architecture (image-based, 4-vectors based, or theory-motivated-variables-based), different models might be learning different discriminating factors, thus combining these learnt representations might help gain better performance.

Although these studies are carried on supervised classification problems , they may as well (or even better) be applied to unsupervised learning cases. 
Beside the results presented here, there are more interesting points to investigate. 
One important topic is the robustness of deep learning models. It's possible that saliency analysis will help us find better ways approaching robust machine learning models.
We hope this effort will develop further into more powerful practical use for physics structure detection in the future.

\subsubsection*{Acknowledgments}

This work is funded by IVADO Fundamental Research Grant. The author would like to thank Gilles Louppe for collaboration in the very early stage. And the author is also grateful for helpful discussions with MILA colleagues. Finally, a thank you goes to Prof. Jean-Francois Arguin for kind support during this work was done.

\small

\bibliography{ref}
\bibliographystyle{plain}

\clearpage
\appendix

\section{Datasets and Neural Net Architecture} 
\label{app:data}
\paragraph{Datasets}
All the jet samples are generated by Pythia8 \cite{Sjostrand:2014zea} and passed to Delphes \cite{deFavereau:2013fsa} for fast detector simulation.
Jets have $p_T \in$ [500, 700] GeV.
They are clustered using anti-kt algorithm \cite{Cacciari:2008gp} with FastJet \cite{Cacciari:2011ma}. Jet clustering cone size is set to be $R=0.8$. 

\paragraph{Neural Net Settings}
The neural networks used in this work is similar as in \cite{Louppe:2017ipp, Cheng:2017rdo}. We embed jets along with their clustering histories which are used as scaffold for building the recursive embedding process. Input features ($p_i, \eta_i, \phi_i, E_i, E_i/E_J, p_{Ti},  \theta_i=2 \arctan(\exp{(-\eta_i)})$) for constituent particles within jets are taken as direct input fed into the neural networks. We use anti-kt clustered jets, then recluster using Cambridge-Aachen algorithm for simplicity in Lund Plane analysis. The network architecture is (RecNN Embedding ($n_{\rm embedding}=70$) $\to$ Dense(70) $\to$ Dense(70) $\to$ Sigmoid).

\section{Transferability of RecNN}
\label{app:transfer}

In the study of \emph{Embedding Transferability}, we use the embedding layer learnt
from classification task A to obtain jet embeddings for task B, then feed these transferred embeddings to a dense network for classification. (i.e. freeze the embedding layer which is initialized with parameters learnt in task A).
In Table \ref{tab:transfer}, we show cross-task transfer AUCs for all the combinations, with
\emph{Base AUC} denoting original full training AUC for B, and \emph{Transfer AUC} denoting 
results from the transferred embeddings.

\begin{table}[htb!]
    \centering
    \begin{tabular}{c|c|c} 
    \toprule
        Tasks &  Base AUC & Transfer AUC \\
        \midrule
        W/QCD $\to$ Top/QCD &  0.926 & \textbf{0.891} \\
        g/q $\to$ Top/QCD & 0.926 & 0.791 \\
        Top/QCD $\to$ W/QCD & 0.957 & \textbf{0.911} \\
        q/g $\to$ W/QCD & 0.957 & 0.822 \\
        W/QCD $\to$ q/g  & 0.861 & 0.763 \\
        Top/QCD $\to$ q/g & 0.861 & 0.759  \\
    \bottomrule
    \end{tabular}
    \caption{Transferability results shown here. In \emph{Base AUC}, the original trained AUC for the target task is shown, while in resulting in  \emph{Transfer AUC}, transferred embedding is used for training the classifier.}
    \label{tab:transfer}
\end{table}

\section{Linear Probing in Other Models}
\label{app:models}

To quickly explore the latent space of other popular architectures, we take a few low-level feature based models which have been used for jet tagging: FCN, LSTM, CNN.

\begin{itemize}
    \item FCN: taking in four momenta of first 30 jet constituents in a $p_T$-ordered manner. The architecture is similar as in \cite{Pearkes:2017hku}. The dense hidden layers have 300, 102, 12, 6 nodes with ReLU activation. Output layer has a sigmoid activation. 
    \item LSTM: taking in four momenta of first 30 jet constituents in a $p_T$-ordered manner. LSTM layer outputs a 70 dimensional embedding, which then is fed into two ReLU dense layers with 50, 20 nodes. Output layer has a sigmoid activation.
    \item CNN: taking in 37$\times$37 grey scale images on ($\eta, \phi$) plane with $p_T$ deposit as pixel intensity.  We take a similar architecture as in \cite{Kasieczka:2019dbj}: (Conv2D*16 $\to$ Conv2D*16 $\to$ MaxPooling2D $\to$ Conv2D*8 $\to$ Conv2D*8 $\to$ MaxPooling2D $\to$ Flatten $\to$ Dense(128) $\to$ Dense(64) $\to$ Sigmoid).
\end{itemize}

\begin{figure}[htb!]
    \centering
    \includegraphics[width=0.24\textwidth]{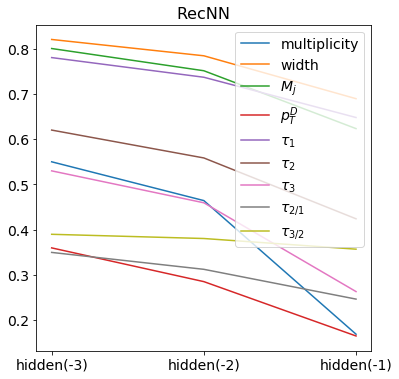}
    \includegraphics[width=0.24\textwidth]{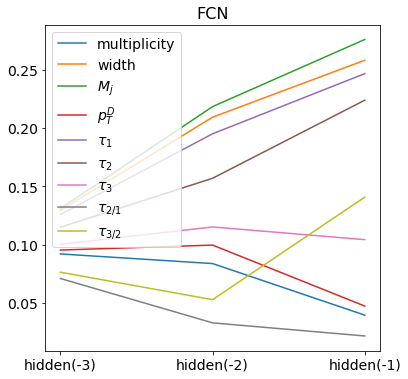}
    \includegraphics[width=0.24\textwidth]{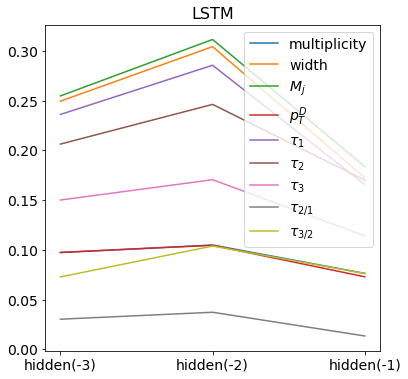}
    \includegraphics[width=0.24\textwidth]{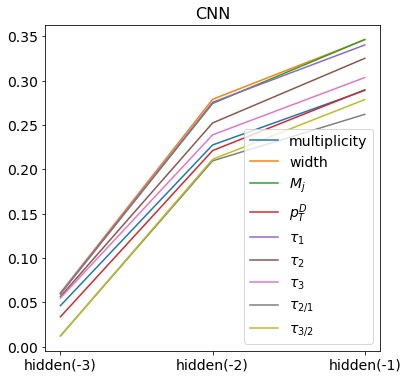} 
    \caption{R2 scores for latest hidden layers for RecNN, FCN, LSTM, CNN. \emph{Hidden(-1)} is the latest layer, \emph{Hidden(-2)} is the second latest, and \emph{Hidden(-3)} is the third latest. For RecNN and LSTM \emph{Hidden(-3)} is just the embedding layer. For CNN, \emph{Hidden(-3)} is the flatten layer.}
    \label{fig:r2_models}
\end{figure}

For simplicity, we only take Top/QCD classification as an instance here. Models are trained on datasets provided in \cite{Butter:2017cot}. R2 scores are shown for 3 latest layers in Fig. \ref{fig:r2_models}.

\section{Extra Plots for Saliency Lund Plane}
\label{app:lund}

\begin{figure}[h]
    \centering
    \includegraphics[width=0.3\textwidth]{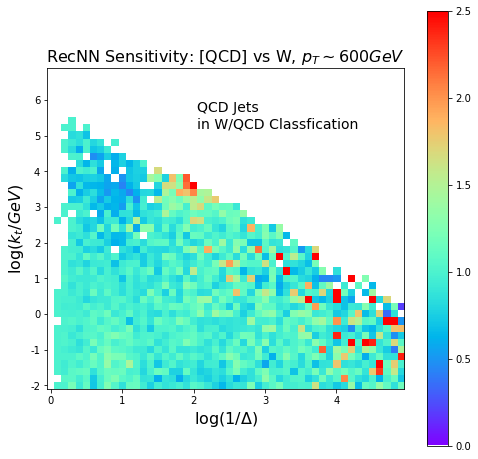}
    \includegraphics[width=0.3\textwidth]{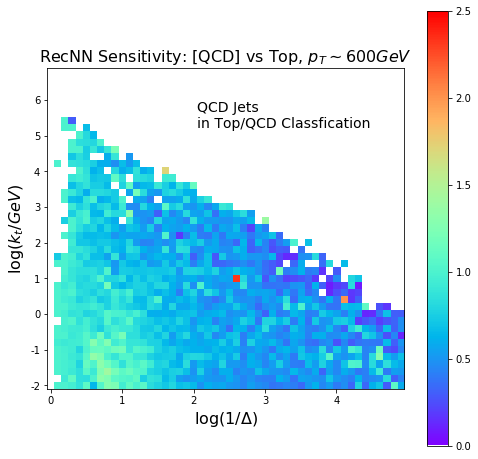}
    \includegraphics[width=0.3\textwidth]{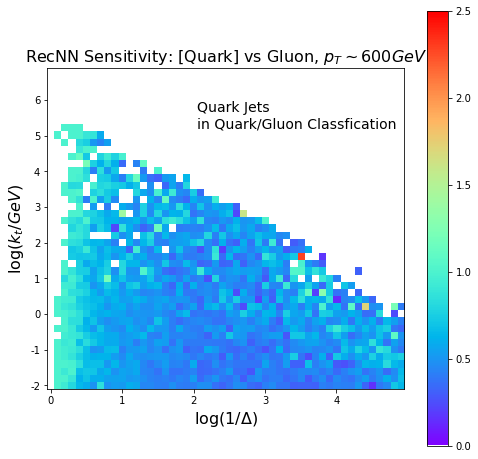} \\
    
    \includegraphics[width=0.3\textwidth]{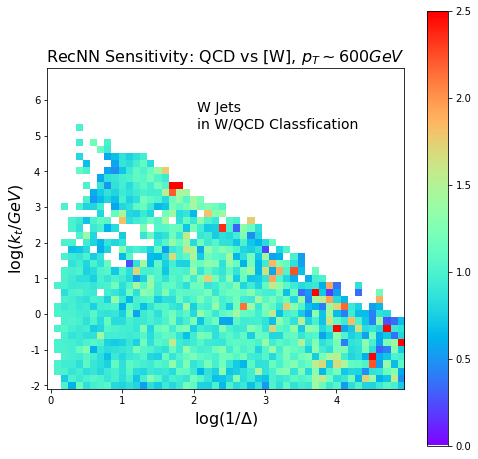}
    \includegraphics[width=0.3\textwidth]{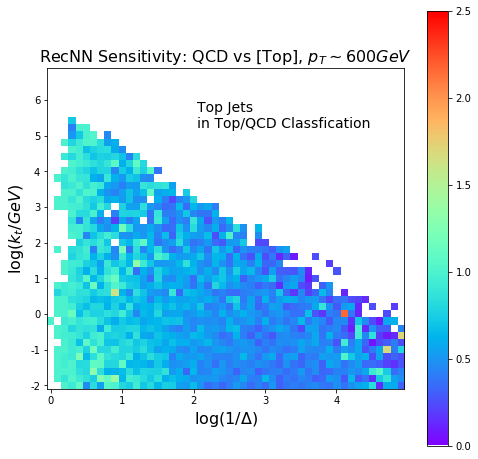}
    \includegraphics[width=0.3\textwidth]{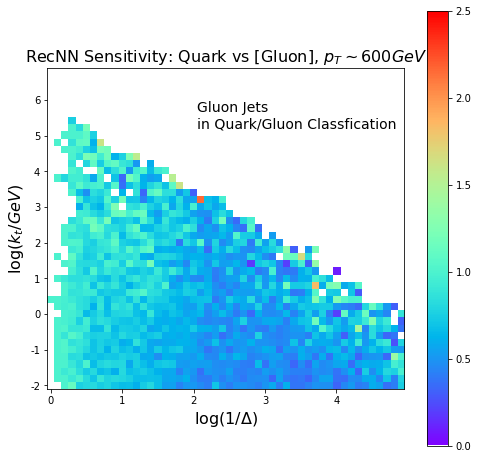}
    \caption{Saliency sensitivity mapped onto Jet Lund Plane. \textbf{Left}: QCD and W jets in W/QCD Classification; \textbf{Middle}: QCD and Top jets in Top/QCD Classification; \textbf{Right}: Quark and Gluon jets in Quark/Gluon Classification.}
    \label{fig:saliencyp}
\end{figure}

\end{document}